\documentclass[pra, 12 pt]{revtex4}
\usepackage[english]{babel}
\usepackage{amsmath}
\usepackage{epsfig}

\begin{document}


\title{\bf Collective excitations and the gap in spectrum of the degenerated Bose gas
}
\author{V.B. Bobrov, S.A. Trigger}
\address{Joint\, Institute\, for\, High\, Temperatures, Russian\, Academy\,
of\, Sciences, 13/19, Izhorskaia Str., Moscow\, 125412, Russia;\\
email:\,satron@mail.ru}

\begin{abstract}
Model of the degenerated weakly non-ideal Bose gas is considered without suggestion on C-number representation of the creation and annihilation operators with zero momentum. The "density-density" correlation function and the one-particle Green function are calculated on basis of the suggestion about C-number representation of the operator of particle density in the Bose condensate. It is shown, that the pole in the "density-density" Green function determines the Bogolyubov's spectrum of the collective excitations, which have the phonon-roton form. At the same time the spectrum of the one-particle excitations possesses a gap, whose value is connected with the density of particles in the Bose condensate.\\

PACS number(s): 05.30.Jp, 03.75.Kk, 03.75.Nt, 05.70.Fh\\

\end{abstract}

\maketitle

\section{Introduction}

The fundamental ideas and approach on explanation of superfluidity in HeII have been formulated by Landau [1,2]. The crucial role in these papers play the statement about the phonon-roton spectrum of
the elementary excitations and the Landau condition for superfluidity. The microscopic model of the weakly non-ideal Bose gas below the condensation temperature $T_0$ has been constructed by Bogolyubov [3] by use the concept of the C-number character of the creation $a_0^+$ and annihilation $a_0$ operators and diaganalization of the approximate Hamiltonian, executed on basis of this concept.

The spectrum of the elementary excitations $\hbar \omega(q)$ in this model corresponds for a small $q$ to the real phonon-roton spectrum in HeII at $T\ll T_0$, which follows from the neutron scattering experiments [4,5]. For large values of $q$ this spectrum corresponds to one for the free particles $\varepsilon_q=\hbar^2 q^2/2m$. The respective diagram technic for the Bose systems at $T<T_0$  has been developed by Belyaev [6].
It is necessary to stress that in the early papers Landau and also Bogolyubov mentioned possibility of the existence of the spectrum with a gap, but later they omitted this idea, because contradiction one with the observations of the phonon branch of spectrum at small $q$ in the neutron scattering experiments. Recently the arguments and the preliminary estimations on coexistence of two branches of spectrum - the Bogolyubov's collective, and a new - one-particle with a gap, have been done in [7-10]. The proof of this concept can be done by the various approach, including the dielectric formalism, developed for the Bose system at $T<T_0$ in [11-13]. In the present paper this statement is justified in the framework of the hypothesis about C-number behavior of the density of the particles in the state with $p=0$.

\section{Density-density correlation function and modified C-number approach}

Let us consider the weakly non-ideal Bose gas at $T<T_0$. Then, the distribution function in momentum space
\begin{eqnarray}
f(p)=\langle a_p^{+}a_p \rangle \label{A1}
\end{eqnarray}
by analogy with the case of an ideal Bose gas takes the form
\begin{eqnarray}
f(p)=\langle N_0 \rangle \delta_{p,0}+f_p^T (1-\delta_{p,0}),  \label{A2}
\end{eqnarray}
Here $a_p^{+}$ ($a_p$) are the operators of creation (annihilation) of a particle with the momentum $\hbar p$, the angle brackets mean averaging on the grand canonical ensemble with the Hamiltonian $H$
\begin{eqnarray}
H=\sum_p \varepsilon_p a_p^{+}a_p + \frac {1}{2V}\sum_{q,\, p_1,\, p_2} u(q)a_{p_1-q/2}^{+} a^{+}_{p_2+q/2} a_{p_2-q/2} a_{p_1+q/2} \label{A3}
\end{eqnarray}
Here $\mu$ is the chemical potential, the function $u(q)$ is the Fourier-component of the interaction potential between the particles, $N=\sum_p a_p^{+} a_p$, $N=a_0^{+} a_0$ - the operator of the number of particles with the momentum equal zero - (the "condensate"),  $f_p^T=\langle a_p^{+}a_p\rangle$  - the one-particle distribution function for the particles with the non-zero momentum (the "over-condensed" particles).

Let us consider the temperature "density-density" Green function $\chi(q,i\Omega)$ ($q\neq 0$)
\begin{eqnarray}
\chi(q,i\Omega)=\frac{1}{V}\langle\langle \rho_q \mid \rho_{-q} \rangle\rangle_{i\Omega} \label{A4}
\end{eqnarray}
 where $\rho_q = \Sigma_p \, a^{+}_{{\bf p-q}/2}a_{{\bf p+q}/2}$  is the Fourier-component of the particle density operator. Then the function $\chi(q,i\Omega)$ can be represented in the form
\begin{eqnarray}
\chi(q,i\Omega) = \frac{1}{V} \,\Sigma_p \,F({\bf p,q},i\Omega),\,\;F(\bf{p,q},i\Omega)=\langle\langle \, a^{+}_{{\bf p-q}/2}a_{{\bf p+q}/2} \mid \rho_{-q} \rangle\rangle_{i\Omega} \label{A5}
\end{eqnarray}
The equation of motion for the function $F({\bf p,q},i\Omega)$  with the Hamiltonian $H$ (Eq.~(\ref{A3})) reads

\begin{eqnarray}
\left(i\Omega +\varepsilon_{\bf{p-q}/2}-\varepsilon_{\bf{p+q}/2}\right)\,F({\bf p,q},i\Omega)= f_{{\bf p-q}/2}-f_{{\bf p+q}/2}-\nonumber\\
\frac{1}{V}\sum_k u(k)\sum_{p_1}\langle\langle \, (a^{+}_{{\bf p+k-q}/2}a^{+}_{{\bf p_1-k}/2}a_{{\bf p_1+k/2}}a_{{\bf p+q}/2}-a^{+}_{{\bf p-q}/2}a^{+}_{{\bf p_1-k}/2}a_{{\bf p_1+k}/2}a_{{\bf p-k+q}/2} )\mid \rho_{-q} \rangle\rangle_{i\Omega} \label{A6}
\end{eqnarray}
Taking into account Eq.~(\ref{A2}), from Eq.~(\ref{A5}) we find that the function $F({\bf p,q},i\Omega)$ has the singularities at ${\bf p}=\pm {\bf q}/2$. Therefore, the density-density function \, $\chi(q,i\Omega)$ can be represented in the form
\begin{eqnarray}
\chi(q,i\Omega) = \frac{1}{V}\,F({\bf q}/2,{\bf q},i\Omega)\,+\frac{1}{V}\,F(-{\bf q}/2,{\bf q},i\Omega)\,+\frac{1}{V}\sum_{p\neq\pm q/2}\,F^T({\bf p,q},i\Omega)\label{A7}
\end{eqnarray}
The index $T$ means, that the respective function concerns to the "over-condensed" particles. Then in the last term in Eq.~(\ref{A7}) we can change summation by integration on momenta.
The functions $F(\pm {\bf q/2,q},i\Omega)$ extracted above satisfy, according to Eqs.~(\ref{A2}),(\ref{A6}), to the equations of motion
\begin{eqnarray}
\left(i\Omega -\varepsilon_q\right)\,F({\bf q}/2,{\bf q},i\Omega)=\langle N_0 \rangle-\frac{1}{V}\sum_{k\neq 0} u(k)\sum_{p_1}\langle\langle \, (a^{+}_{\bf k} a^{+}_{{\bf p_1}-{\bf k}/2}a_{{bf p_1}+{\bf k}/2}a_{\bf q} -\nonumber\\ a^{+}_0 a^{+}_{{\bf p_1}-{\bf k}/2}a_{{\bf p_1}+{\bf k}/2}a_{{\bf q-k}})\mid \rho_{-q} \rangle\rangle_{i\Omega} \label{A8}
\end{eqnarray}
\begin{eqnarray}
\left(i\Omega +\varepsilon_q\right)\,F(-{\bf q}/2,{\bf q},i\Omega)=-\langle N_0 \rangle-\frac{1}{V}\sum_{k\neq 0} u(k)\sum_{p_1}\langle\langle \, (a^{+}_{{\bf k-q}} a^{+}_{{\bf p_1}-{\bf k}/2}a_{{\bf p_1}+{\bf k}/2}a_0 -\nonumber\\ a^{+}_{-{\bf q}} a^{+}_{{\bf p_1}-{\bf k}/2}a_{{\bf p_1}+{\bf k}/2}a_{-{\bf k}})\mid \rho_{-q} \rangle\rangle_{i\Omega}. \label{A9}
\end{eqnarray}
Let us consider further the case of strongly degenerated gas, where $T\rightarrow 0$. In this case we accept $\langle N_0\rangle \rightarrow \langle N\rangle$.
According to the Bogolyubov's procedure, let us extract in the right sides of Eqs.~(\ref{A8}),(\ref{A9}) the main terms, which are determined by the maximum quantity of the operators $a^{+}_0$ and $a_0$. In the limit of a strong degeneration and a weak interaction the other terms can be omitted. Then for ${\bf q}\neq 0$  Eqs.~(\ref{A8}),(\ref{A9}) read
\begin{eqnarray}
\left(i\Omega \mp \varepsilon_q\right)\,F(\pm {\bf q}/2,{\bf q},i\Omega)=\pm \left(\langle N_0 \rangle+\frac{1}{V} u(q) \langle\langle \, (a^{+}_0 a^{+}_0 a_{\bf q} a_0 +a^{+}_0 a^{+}_{-{\bf q}}a_0 a_0)\mid \rho_{-q} \rangle\rangle_{i\Omega}\right) \label{A10}
\end{eqnarray}

Let us suppose that the operator of the quantity of particles in the "condensate"  $N_0$ is the C-number ($N_0=<N_0>$). At the same time the operators  $a^{+}_0$ è $a_0$  are not the C-numbers ($a_0 a^{+}_0-a^{+}_0 a_0=1$) in contrast with the theory of Bogolyubov.  In this case from Eqs.~(\ref{A10}) straightly follows
\begin{eqnarray}
\left(i\Omega \mp \varepsilon_q\right)\,F(\pm {\bf q}/2,{\bf q},i\Omega)=\pm \left(\langle N_0 \rangle+\frac{N_0}{V} u(q) \{F({\bf q}/2,{\bf q},i\Omega)+F(-{\bf q}/2,{\bf q},i\Omega\}\right) \label{A11}
\end{eqnarray}
From Eqs.~(\ref{A11}) one find the solutions for the functions $F(q/2,q,i\Omega)$ and $F(-q/2,q,i\Omega)$                                                                                             \begin{eqnarray}
F({\bf q}/2,{\bf q},i\Omega)=\frac{<N_0>(i\Omega+\varepsilon_q)}{(i\Omega)^2-(\hbar\omega(q))^2};\;\; F(-{\bf q}/2,{\bf q},i\Omega)=-\frac{<N_0>(i\Omega-\varepsilon_q)}{(i\Omega)^2-(\hbar\omega(q))^2}\label{A12}
\end{eqnarray}
\begin{eqnarray}
\hbar\omega(q)\equiv \sqrt {\varepsilon_q^2+2 n_0 u(q) \varepsilon_q}\label{A13}
\end{eqnarray}
where $n_0\equiv\langle N_0\rangle/V$ is the average density of particles in the "condensate". The relation (\ref{A13}) for the spectrum exactly corresponds to the known Bogolyubov's expression.
Inserting (\ref{A12}) in (\ref{A11}) and taking into account that for a strong degeneration and a weak interaction the contribution of the functions $F^T({\bf p,q},i\Omega)$ is negligible, we arrive at the expression for the "density-density" Green function [11,12]
\begin{eqnarray}
\chi(q,i\Omega) = \frac{2 n_0 \varepsilon_q}{(i\Omega)^2-(\hbar\omega(q))^2}\label{A14}
\end{eqnarray}
The function $\chi(q,i\Omega)$ can be continued analytically to the upper semi-plane of the complex $z$, where this function coincides with the retarded "density-density" Green function $\chi^R(q,z)$ (with the change $i\Omega$ on $\hbar z$). As it is well known, the singularities of the function $\chi^R(q,z)$ determine the spectrum of collective excitations in the system.
Therefore, in the suggestion about C-number behavior of the operator $N_0$ we reproduce the Bogolyubov's result for the spectrum of excitations in the degenerated and weakly interacted Bose gas.

\section{One-particle spectrum with the gap}

The necessary conditions for correctness of the above derivation are the conditions $lim_{T\rightarrow 0}\langle N_0 \rangle=\langle N \rangle$ and finiteness of the function $f_p^T$  for $p \rightarrow 0$. Let us show that these conditions are fulfilled in the degenerated weakly non-ideal Bose gas.
For this purpose we calculate the one-particle Green function $g(p, i\Omega)$ for ${\bf p}\neq 0$
\begin{eqnarray}
g(p,i\Omega) = \langle\langle \, a^{+}_{\bf p} \mid a_{\bf p} \rangle\rangle_{i\Omega}\label{A15}
\end{eqnarray}

The equation of motion for the Green function $g(p,i\Omega)$ for $p\neq 0$ reads
\begin{eqnarray}
\left(i\Omega - \varepsilon_p +\mu\right)\,g(p,i\Omega)=1+\frac{1}{V}\sum_{k} u(k)\sum_{p_1}\langle\langle \, a^{+}_{{\bf p_1+k}} a_{{bf p_1}}a_{{\bf p+k}}\mid a^{+}_{\bf p} \rangle\rangle_{i\Omega}=\nonumber\\1+\frac{<N>}{V} u(0)g(p, i\Omega)+\frac{1}{V}\sum_{k\neq 0} u(k)\sum_{p_1}\langle\langle \, a^{+}_{{\bf p_1+k}} a_{{\bf p_1}}a_{{\bf p+k}}\mid a^{+}_{\bf p} \rangle\rangle_{i\Omega}. \label{A16}
\end{eqnarray}
where $u(0)=u(p=0)=lim _{p\rightarrow 0}u(p)$.
As for the "density-density" response function, we consider the case of a strong degeneration and a weak interaction and extract in the right side of Eq.~(\ref{A16}) the terms with maximum quantity of the operators $a^{+}_0$ and $a_0$.  Then from Eq.~(\ref{A16}) one find
\begin{eqnarray}
\left(i\Omega - \varepsilon_p +\mu^{(0)}\right)\,g(p,i\Omega)=1+\frac{1}{V} u(p)\langle\langle \, a^{+}_0 a_{\bf p} a_0\mid a^{+}_{\bf p} \rangle\rangle_{i\Omega}(1-\delta_{p,\,0}), \label{A17}
\end{eqnarray}
where $\mu^{(0)}=\mu-nu(0)$ and $n=\langle N \rangle /V$ is the average density of the particles. Further, suggesting again the C-number behavior of the operator $N_0$ and taking into account that in the approximation under consideration $\mu^{(0)}=0$, we find for ${\bf p}\neq 0$
\begin{eqnarray}
g(p,i\Omega)=\frac{1}{i\Omega -E_p}, \label{A18}
\end{eqnarray}
The expression for the spectrum $E_p$ is equal
\begin{eqnarray}
E_p=\varepsilon_p+n_0 u (p), \label{A19}
\end{eqnarray}
From Eqs.~(\ref{A18}),(\ref{A19}) for $T \ll T_0$ directly follows
\begin{eqnarray}
f_p^T=\frac{1}{\exp(\beta E_p)-1}, \label{A20},
\end{eqnarray}
where $\beta=1/T$.
Therefore, the function $f_p^T$ is finite for $p \rightarrow 0$. Moreover,
\begin{eqnarray}
f_p^T \rightarrow 0, \label{A21}
\end{eqnarray}
for the limit $p\rightarrow 0$, $T \rightarrow 0$ in opposite to the case of the ideal gas, where $f_p^{T, id} \rightarrow \infty$ for $p\rightarrow 0$. Accordingly, the relation (\ref{A2}) is valid for the interacted system.
Besides that, in the one-particle excitation spectrum appears the gap $\Delta=E_{p \rightarrow 0}=n_0 u(0)$.

\section{Conclusions}

Therefore, even a weak interaction between Bose particles leads to the drastic distinction with the case of the ideal Bose gas. This distinction takes place not only for the collective excitations, which are described by the "density-density" correlation function, but also in the distribution function and the one-particle excitations of the over-condensed particles.
In the case under consideration, for strong degeneration $T \ll T_0$, the influence of the over-condensed particles and the temperature effects are small. However, the conclusion about existence of the gap is evidently valid in the region of temperatures $T \leq T_0$. In this connection we have underline that the spectrum $E_p$ (\ref{A19}) satisfies the Landau criterion of superfluidity for transitions between the condensed and over-condensed particles. For the transitions between the over-condensed particles this criterion violates, as it was mentioned by Bogolyubov.

\section*{Acknowledgment}

The authors are thankful to the Netherlands Organization for Scientific Research (NWO) for
support of this work in the framework of the grant ¹ 047.017.2006.007.

\end{document}